# CHEmical-shift selective Adiabatic Pulse (CHEAP): Fast and High Resolution Downfield 3D [1]H-MRSI at 7T

Guodong Weng[# 1,2], Piotr Radojewski[# 1,2], Sulaiman Sheriff[3], Arsany Hakim[1], Irena Zubak[4], Johannes Kaesmacher[1], Johannes Slotboom[1,2]

[#] Equally contributing first authors

Johannes Slotboom: johannes.slotboom@insel.ch

Guodong Weng (corresponding author): guodong.weng@unibe.ch

Affiliations:

1. Institute for Diagnostic and Interventional Neuroradiology, Inselspital, University Hospital and University of Bern, Switzerland
2. Translational Imaging Center, sitem-insel, Bern, Switzerland
3. University of Miami, Miami, FL, United States
4. Department of Neurosurgery, Inselspital, University Hospital and University of Bern, Switzerland

## Abstract

The key molecules such as triphosphate (ATP), glutathione (GSH), and homocarnosine (hCs) - central to metabolic processes in the human brain remain elusive or challenging to detect with upfield [1]H-MRSI. Traditional 3D [1]H-MRSI *in vivo* faces challenges, including a low signal-to-noise ratio and magnetization transfer effects with water, leading to prolonged measurement times and reduced resolution. To address these limitations, we propose a downfield 3D-MRSI method aimed at measuring downfield metabolites with enhanced spatial resolution, and speed acceptable for clinical practice at 7T. The **CHE**mical-shift selective **A**diabatic **P**ulse (CHEAP) technique was integrated into echo-planar spectroscopic imaging (EPSI) readout sequence for downfield metabolite and water reference 3D-MRSI. Five healthy subjects and two glioma patients were scanned to test the feasibility. In this work, CHEAP-EPSI technique is shown to significantly enhance spatial the resolution to 0.37 ml while simultaneously reducing the scan time to 10.5 minutes. Its distinct advantages include low specific absorption rate, effective suppression of water and lipid signals, and minimal baseline distortions, making it a valuable tool for research or potentially diagnostic purposes. CHEAP-EPSI improves the detection sensitivity of downfield metabolites like N-acetyl-aspartate (NAA+) and DF8.18 (ATP&GSH+), and offers new possibilities for the study of metabolism in healthy and diseased brain.

## 1. Introduction

Over the past four decades, localized *in vivo* [1]H-MRS(I) of the brain has evolved into a powerful tool for investigating not only normal brain metabolism, but also aging [1,2], and various pathologies, including brain tumors [3,4], hypoxic and ischemic conditions [5], and inborn diseases [6]. The majority of MRS studies performed in the past have focused on the upfield part of the [1]H-spectrum, encompassing the chemical shift offset range between 0.00 ppm and 4.67 ppm. In contrast, there has been considerably less exploration of the downfield part of the spectrum [7], where the chemical shift offset range is higher than 4.67 ppm, especially *in vivo* [1]H-spectra of human tissues [8–11]. A few pioneering studies in downfield MRS were published in the 1980-1990s [12–15]. Despite continuous growth, downfield [1]H-MRS research has been notably less prolific compared to upfield [1]H-MRS. Notable contributions emerged in the 1980s, including studies on phenylalanine [11,16–18], α-glucose [19], homocarnosine [10], histidine [9,13], among others. Many important metabolites that are present in the downfield 1H-MRS range (<4.67 ppm) but cannot or can hardly be detected in the upfield [1]H-MRS range (>4.67 ppm), for instance adenosine triphosphate (ATP), homocarnosine (hCs), histidine (His), etc. GSH and ATP are regarded as key metabolites with abnormalities described brain tumors [20], stroke [21,22] and Parkinson's disease [23].

In more recent literature (Fichtner et al. [24]), potential factors contributing to the limited number of downfield [1]H-MRS studies are elucidated. These factors include: **1.** Low Signal-to-Noise Ratio and related challenges in detecting specific peaks due to weak signals; **2.** Non assigned peaks and overlapping peaks: Difficulty distinguishing individual substances due to complex overlaps and knowledge to which metabolites/substances these peaks should be attributed; **3.** In downfield proton magnetic resonance spectroscopy (MRS), complex relaxation mechanisms encompass the interplay of $T_1$ and $T_2$-relaxation, chemical exchange (magnetization transfer), and J-coupling effects. **4.** Wide range of $T_1$ values: variations in $T_1$ values add complexity to resonance characterization; **5.** Large measurement volumes often associated with inhomogeneously broadened components; **6.** Technical challenges in quantification due to the factors mentioned above.

Due to challenges described above and the limited number of studies the role that downfield MRS can play in research or diagnostic setting is still undefined.

Owing to the challenges of low Signal-to-Noise Ratio (SNR) (issue **1**) and water saturation (issue **3**), previous studies have primarily limited the investigations to single-voxel volume MRS. Nonetheless, recent advancements have expanded the horizons of downfield single-voxel volume MRS to 2D [25] and 3D MRSI [26] using spectral-spatial excitation at 3T. Notably, a whole-brain 3D MRSI pilot study examining brain tumor patients has underscored potential value of downfield MRSI in clinical research and even diagnostic routine [27]. This pioneering work achieved a nominal resolution of 0.74 ml with an acquisition time (TA) of approximately 22 minutes for metabolite MRSI, and a further 11 minutes for water reference MRSI (total TA = ~33 minutes). To our best knowledge, the first downfield 3D-MRSI study at 7T was published in 2020 using BURP (Band-selective, Uniform Response, Pure-phase) pulses and Chemical Shift Imaging (CSI) recording [28] with low resolution (27 ml). In 2023, high resolution MRSI (0.34 ml) at 7T was achieved using CHEmical-shift selective Adiabatic Pulse (CHEAP) together with EPSI recoding [29], while the use of asymmetric adiabatic pulse was first developed in 1996 [30]. In 2024, the above-mentioned approach using spectral-spatial excitation with CSI recoding at 7T was published [31], which resulted in a nominal resolution of 0.74 ml with about 10 minutes for metabolite MRSI and another ~8 minutes for water reference MRSI (total TA = ~18 minutes).

The purpose of this study was to develop a dedicated pulse sequence and to perform measurements of downfield [1]H-brain spectra between the 6.2–9.2 ppm range, focusing on achieving higher resolution and reducing acquisition time with ultra-high field MR (7T). To accomplish this, we use an adapted version [29,32,33] of the echo-planar spectroscopic imaging (EPSI) [34] sequence, employing the novel approach CHEAP, distinct from prior methodologies [25–27].

We hypothesized that this novel approach could simultaneously address issues such as low SNR of downfield metabolites (issue **1**), avoids saturating the water signal (issue **3**), consequently improving the SNR of downfield metabolites, and allowing for substantially smaller acquisition volumes, enabling multi-slice measurements (issue **5**). The method is a low Specific Absorption Rate (SAR) technique, implicit water and lipid suppression, minimal baseline artifacts, facilitating straightforward peak integration maps [32].

The primary aim of the present proof-of-concept study is to test these hypotheses in first *in vivo* measurements.

## 2. Methods

### 2.1. Scanner

All MRI and MRSI acquisitions were performed on a Siemens 7T MR scanner in clinical mode (MAGNETOM Terra, Germany, software version VE12U-SP01) using the Nova 1Tx 32Rx head coil.

### 2.2. Asymmetric Chemical-shift selective Adiabatic Pulse (CHEAP)

The asymmetric CHEAP shape was designed by complex secant hyperbolic function (see **Figure 1A**) with two different parameter-sets: (1.) the increasing part of the pulse; (2.) the decreasing part of the pulse:

$$B_1^+(t) = \Omega_0 \cdot \mathrm{sech}(\beta t)^{1+\mu i}$$

Where $B_1^+$ is the RF, $\Omega_0$ is the maximum amplitude of the RF pulse.

1. The first half (from 0 to 4 ms) of the asymmetric CHEAP has following parameters: pulse duration = 8 ms, $\Omega_0$ = 989 Hz, $\beta$ = 771, and $\mu$ = 1.96.
2. The second half (from 4 to 10 ms) of the asymmetric CHEAP has following parameters: pulse duration = 12 ms, $\Omega_0$ = 989 Hz, $\beta$ = 1154, and $\mu$ = 4.40.

This designed CHEAP has a bandwidth of 900 Hz with a maximum of 989 Hz $B_1^+$ is 100% above the adiabatic threshold.

### 2.3. Sequence design

As is shown in **Supplementary Figure S1**, the sequence was developed based on EPSI-sequence [34] using the Siemens IDEA-VE12U-SP01 programming environment and it consists of the following parts:

1. Excitation: A slice-selective sinc-Gaussian pulse with 6 ms duration, 5.5 kHz BW, and a 65-degree Ernst flip angle is used for excitation.
2. Refocusing: The CHEAP pair is used for both refocusing, which is described above. The spoiler gradient pairs are placed directly adjacent to the CHEAP pair. The total gradient durations are 2.5-2.5-3.6-3.6 ms and having following amplitudes of 14.7-14.7-28.0-28.0 mT/m and 14.7-14.7-3.1-3.1 mT/m in X- and Z-axes, respectively.
3. EPSI readout for metabolite [34]: The readout is composed by 1024 gradient lobes which generates 512 even and odd echoes. The ramp time, total duration, and amplitude of each gradient lobe are set to be 160 μs, 390 μs, and 19.92 mT/m, respectively. This is followed by a spoiler gradient with 20 ms duration and 5 mT/m amplitude. The carrier frequency was set to be 7.67 ppm instead of 4.7 ppm of water. The aquired spectral bandwidth is from 5.51 to 9.83 ppm.

4.  Excitation for water reference: The same slice-selective sinc-Gaussian pulse as in part 1 is used, but with a flip angle of 10 degrees

5.  EPSI-readout for water reference [34]: The carrier frequency was set to 4.7 ppm water and the other parameters are the same as the readout scheme in part 3.

## 2.4. Sequence

The CHEAP-EPSI sequence [32,34] (see **Figure 1A**) with asymmetric CHEmical-shift selective Adiabatic Pulse (CHEAP) [35] was applied with following parameters:

TE = 40 ms, TR = 1300 ms, nominal matrix = $65 \times 23 \times 5$ ($4.3 \times 7.8 \times 11$ mm), zero filling matrix = $65 \times 42 \times 6$ ($4.3 \times 4.3 \times 9.2$ mm), FOV = $280 \times 180 \times 55$ mm, averages = 10, and TA = 21 min. The refocusing CHEmical-shift selective Adiabatic Pulse (CHEAP) is 10 ms duration. The bandwidth (full width at 50% maximum) ranges from $6.2 - 9.2$ ppm (**Figure 1B**). No water and lipid suppression needs to be applied due to the implicit water and lipid suppression of the CHEAP.

## 2.5. Subjects

Subjects 1 to 5 were healthy volunteers, a 59-year-old male (Subject #1), a 58-year-old male (Subject #2), a 31-year-old male (Subject #3), a 28-year-old female (Subject #4), and a 29-year-old female (Subject #5).

Subject #6 was a 47-year-old female patient with IDH-mutated WHO Grade 2 astrocytoma. Subject #7 was a 49-year-old male patient with IDH-mutated Grade 3 astrocytoma. In these two patients, only two averages of CHEAP-EPSI were administered before surgery, with a total acquisition time of 4:12 min.

## 2.6. Data reconstruction and pre-post-processing

For the data reconstruction and pre-post-processing, the Metabolic Imaging Data Analysis System (MIDAS)[36], and MATLAB R2019b were used. Further details are provided in the supplementary material (**Supplementary Figure S2**).

## 2.7. Mapping

The DF (DownField) 7.82 (7.66-7.97 ppm; NAA+; **Figure 1G**), and DF 8.18 (8.05-8.33 ppm; ATP&GSH+; **Figure 1G**) maps were generated by absorption peak integration (**Figure 2**, **3**, **4**, and **6**). The metabolites signal was voxel wise divided by water reference signal to correct $B_1^+$-inhomogeneities of the non-adiabatic excitation pulse and $B_1^-$-inhomogeneities of the coils used for detection. A 2D linear interpolation was applied for each slice, transferring the matrix size from 65×42 to 1025×657.

## 2.8. SNR

The SNR for each voxel of the 10.5-minute acquisition is calculated according to the following equation:

$$SNR = \frac{NAA+_{max}}{SD_{noise}}$$

Where $NAA+_{max}$ is the maximum of the real part of the NAA+ peak, and $SD_{noise}$ is the standard deviation of the complex value of the first 200 points (8.99-9.83 ppm) of the spectrum.

## 2.9. Deviation

The deviation ($Dev$) of NAA+ value for each voxel between the acquisition of 10.5 and 21 minutes is calculated according to the following equation:

$$Dev = \frac{A_{NAA+\_10.5} - A_{NAA+\_21}}{A_{NAA+\_21}}$$

Where $A_{NAA+\_10.5}$ and $A_{NAA+\_21}$ are the peak integration of NAA+ for the acquisition of 10.5 and 21 minutes respectively. In **Figure 5B**, the deviation map shows the absolute value $|Dev|$, while the bar plot (**Figure 5D**) shows the real value $Dev$.

## 3. Results

*In vivo* measurements for downfield metabolites mapping were performed in 5 healthy individuals (29-59 years old, three males and two females) using CHEAP-EPSI (sequence #1) covering a large 3D brain slab as well as in 2 glioma patients using the same protocol.

The downfield spectroscopic recordings revealed a large number of metabolite and macromolecule contributions, encompassing both identified and unidentified substances. **Figure 1D-G** presents spectral data acquired from two healthy subjects, denoted as subjects #1 and #3. The spectra exhibit two discernible peaks at 7.82 ppm and 8.18 ppm. The peak at 7.82 ppm is primarily associated with N-acetyl aspartate (NAA), and amide group of proteins, referred to as NAA+ in this work. In contrast, the peak in the vicinity of 8.18 ppm (8.05-8.33 ppm) manifests increased complexity (**Figure 1D,F,G**), to which many resonances contribute. According to literature, there are a few resonances around this range: ATP (8.22 ppm), GSH (8.18 ppm), homocarnosine (hCs, 8.08 ppm), nicotinamide adenine dinucleotide (NAD+, 8.18 and 8.21 ppm), N‐acetyl aspartyl glutamate (NAAG, 8.26 ppm), and amide group of proteins [37]. The peak at ~8.18 ppm was referred to as ATP&GSH+, as ATP and GSH have the highest concentration (approximately 2-4 mmol) among the metabolites mentioned above [37–39].

In our study, 3D peak integration maps of the ATP&GSH+ (8.05-8.33 ppm), NAA+ (7.66-7.97 ppm), as well as the ATP&GSH+ to NAA+ ratio were generated in healthy subject #1-5. **Figure 2** shows three distinct slices of each subject. Evidently, this ATP&GSH+ peak exhibits increased levels in white matter regions, while registering reduced levels in gray matter regions. In contrast, the NAA+ peak displays a relatively subdued contrast in its distribution between gray and white matter regions. Additionally, the ATP&GSH+ to NAA+ ratio highlights elevated levels within white matter regions.

**Figure 3-4** shows all the slices of ATP&GSH+ and NAA+ map for subject #1 with different average numbers (total acquisition time). It is noteworthy that the spectroscopic image quality of the mapping is achieved by an average number of 5 signal averages (total acquisition time of 10.5 minutes) is very similar to that achieved by an average number of 10 (total acquisition time of 21 minutes).

Quantitative analysis of SNR and deviation in each voxel was performed for all five healthy subjects (#1-5). **Figure 5B** shows the SNR of NAA+ maps with 10.5 minutes TA, and the NAA+ value deviation between the 10.5 minutes and 21 minutes acquisition. The SNRs (median ± SD) were 6.12±1.52, 6.22±1.58, 7.38±1.89, 7.03±1.83, and 6.73±1.85 for subjects #1-5, respectively (**Figure 5C**). In addition, the deviations (median ± SD) were -0.016±0.106, 0.002±0.083, -0.003±0.077, -0.008±0.094, and 0.001±0.086 for subjects #1-5, respectively (**Figure 5D**).

Subsequently, CHEAP-EPSI was applied to two glioma patients (subject #5-6, IDH-mutant astrocytoma, Grade 2 and 3, respectively). As depicted in **Figure 6,** both the ATP&GSH+ (8.18 ppm) and NAA+ (7.82 ppm) peaks exhibit a reduction within the tumor region. The decrease in NAA+ surpasses that in ATP&GSH+, resulting in an elevated ATP&GSH+ to NAA+ ratio within the tumor area.

## 4. Discussion

### 4.1. Without water and lipid suppression

To preserve the signal intensities of numerous downfield metabolites engaged in magnetization exchange with water, [24,40] an important property of CHEAP was utilized in all experiments illustrated across **Figures 1** to **6**. Deviating from conventional practices, we intentionally omitted the application of the water suppression pulse. In this context, the water signal around ~4.67 ppm and the lipid signal spanning approximately 5.2 to 5.3 ppm experience precise dephasing through the application of CHEAP, [32] encompassing a selectively refocused passband spanning 6.2 to 9.2 ppm. This dephasing additionally extends to other signals residing upfield of water ($\leq$ 4.67 ppm), resulting in the emergence of a clean downfield spectrum characterizing 3D-resolved downfield Magnetic Resonance Spectroscopic Imaging (MRSI), combined with the rapid MRSI technique EPSI. [34]

This novel approach not only improves SNR, but also significantly simplifies the experimental sequence, streamlining implementation and application processes, and improves the robustness of the acquisitions. Moreover, the utility of CHEAP-EPSI for downfield MRSI can be effectively transferred to the 3T environment by extending the pulse duration of CHEAP, thereby achieving comparable and sufficient water and lipid dephasing effects.

### 4.2. Downfield spectra and peak assignment

In the spectra shown across **Figure 1D, F, G**, the clear presence of two primary peaks at 7.82 ppm and 8.18 ppm commands attention. Within the established MRS-literature, there is a consensus that N-acetylaspartate (NAA) resonates at 7.82 ppm, while resonances at 8.22, 8.18, and 8.08 ppm can most likely be attributed to ATP, GSH, and hCs, respectively. [37,40] Notably, the resonance frequencies of ATP and hCs remain responsive to pH fluctuations within the physiological range.

Due to the low SNR and the close resonances, it is not possible in this work to distinguish the individual resonances around 8.18 ppm, which may originate from ATP, GSH, hCs, NAD+, NAAG and the amide group of proteins.

In this work, we label 'NAA+' and 'ATP&GSH+' upon the expanded peaks at 7.82 (7.66-7.97) ppm and 8.18 (8.05-8.33) ppm, respectively, indicating that they are merged with contributions from other resonances.

### 4.3. Peak integration maps

The two main peaks, NAA+ and ATP&GSH+, are well separated from each other. In addition, the spectral baseline is quite flat due to the good suppression of water and lipids by CHAP (**Figure 1D, F, G**). It is worth emphasizing that the spectra shown in **Figure 1** cover the entire recorded ppm range (5.5-9.8 ppm), which means that no truncations were made to show only the clean spectral range. Furthermore, no water or lipid removal algorithm and no baseline correction were applied. Therefore, peak integration could be used on the good quality spectra to generate maps for the two peaks in this work.

**Figure 2** shows the NAA+, ATP&GSH+, and ATP&GSH+ to NAA+ ratio maps for *five* healthy subjects, revealing several insights. The ATP&GSH+ level is clearly higher in the white matter compared to its level in the gray matter for all subjects. Maps for all slices (subject #2-5) with color bar were shown in **Supplementary Figure S3-6**.

**Figures 3-4** show the ATP&GSH+ and ATP+ maps for subject #1 at different total acquisition times. Since a ~10-minute acquisition is usually the maximum tolerable for an MR sequence in clinical use, it is of interest to compare the quality of the maps between the TA of 10.5 and 21 minutes. Remarkably, the map shape achieved with 5 averages (total acquisition time of 10.5 minutes) closely resembles that of 10 averages (total acquisition time of 21 minutes), while maintaining a nominal spatial resolution of 0.37 ml.

## 4.4. Quantitative comparisons

To better illustrate the comparison between the two different acquisition times, SNR for NAA+ of 10.5 minutes acquisition and deviation between the two acquisitions (10.5 and 21 minutes) are shown in **Figure 5**. The spectra of two acquisitions (10.5 and 21 minutes) for different voxels in subject #1 were shown in **Supplementary Figure S7**. The median SNR of a single voxel (a nominal spatial resolution of 0.37 ml) in the selected slice for all five subjects is in the range of 6.1-7.4 (**Figure 5B-C**). In comparison, previous work has reported that the average SNR of up to four voxels (a nominal resolution of 0.74 ml) is in the range of 3.4-7.0 for different locations in the brain [31].

The deviation maps show the noise-like pattern (**Figure 5B**), and the median values for all five subjects are less than 2% while the standard deviations are less than 11% (**Figure 5C**). This underpins the comparison in **Figure 3-4** that the downfield maps of 10.5 minutes acquisition are comparable with the 21 minutes acquisition.

In short, our approach doubles the spatial resolution to 0.37 ml and reduces the acquisition time to just 10.5 minutes in downfield MRSI at 7T, leveraging the full potential of our CHEAP-EPSI sequence. Comparatively, the recent work of Ozdemir et al.[31] encompassing whole-brain downfield MRSI at 7T achieves a nominal resolution of 0.74 ml with a total acquisition time (both metabolite and water reference MRSI) of approximately 18 minutes. Our approach reduces acquisition time by almost forty percent to around 10 minutes, while simultaneously doubling spatial resolution, thus suggesting a leap forward in the realm of downfield [1]H-MRSI methodologies.

## 4.5. Glioma patients

Two glioma patients (IDH-mutant, no 1p/19q codeletion, astrocytoma) were examined before surgery in our study, as depicted in **Figure 6**, with an expedited ~4-minute acquisition comprising 2 averages. The maps featuring ATP&GSH+, NAA+, and the ATP&GSH+ to NAA+ ratio clearly delineate the lesion site. Further studies are needed to understand the implications of these downfield resonances within the context of glioma. Moreover, these patients were subjects in a study examining prediction of IDH mutation status through detection of 2HG, encompassing a 9-minute acquisition for upfield MRSI and spectral editing and additional acquisition for downfield MRSI(the corresponding metabolite maps for subject #6 were shown in **Supplementary Figure S8**).[32,41] This underscores the feasibility of achieving high-resolution 3D-resolved MRSI encompassing downfield, upfield, and spectral editing within 13 minutes. The potential impact of comprehensive metabolites examinations extends beyond the glioma research, into diverse conditions such as aging and dementia.

## 4.6. Limitations of the study

The primary limitation of our study is the assignment and quantification of metabolites through spectral fitting. It is important to note, however, that this limitation is not exclusive to our investigation but rather constitutes a pervasive challenge within the field of downfield MRS since a correct and complete prior knowledge spectral basis set which can be used for spectral fitting is yet not available. In recent investigations,[25,26] an approach entailed employing a basis set comprising nine distinct Gaussian peaks for fitting the downfield spectra. Despite this advancement, the issue of unambiguous assignment of metabolites to individual peaks remained unresolved. This highlights the need for further research to refine peak assignments.

There are two reasons why peak integration is used in this study. The first reason is that CHEAP refocusing results in clean $^1$H spectra (two separated peaks and flat baseline) which allows robust and meaningful peak integration. The second reason is the computation time which is considerable for these 3D-MRSI datasets; elsewhere we have shown that peak integration and leads in very similar results.[42] In the meantime we have solved the long computation time hurdle for prior knowledge based NLLS-fitting [43] by at least two orders of magnitude in computation speed.

A promising approach to address this challenge lies in the use of higher magnetic field strengths. Elevating the main magnetic field strength ($B_0$) to a higher field (>7T) presents a potential solution, as the resonance offset of metabolites scales proportionally with the field strength. This property would make nearby metabolites, which often show up as broadened peaks at 7T, differentiable. This approach enables a more precise and differentiated understanding of the composition of metabolites.

Another limitation is the relatively long TE (40 ms) in this study. The minimum TE is limited by the long duration of the CHEAP pulses. The design of the pulse is to refocus the downfield region while dephasing the lipid (~5.3 ppm) and water, therefore a steep transition band between 5.3-7 ppm is required. One can further reduce the pulse duration (i) by compensating for a less steep transition band, as the current stopband is at 5.6 ppm, which is still 0.3 ppm away from lipid or (ii) by making the pulse more asymmetric to maintain the steepness of the transition band. Furthermore, a higher magnetic field (>7T) would allow for a shorter pulse duration because the metabolite resonances in Hz are scaled by the magnetic field. Therefore, this method is expected to work with a shorter TE at an even higher magnetic field.

## Conclusion

Downfield studies are extremely rare compared to upfield MRSI. This work represents a proof-of-concept that supports the hypotheses of our study. Our findings show that the proposed approach effectively dephases water and lipid signals without the need for additional suppression pulses, thereby avoiding saturation of downfield resonances that exhibit magnetization transfer with water resonance. The overall SNR is adequate to achieve high-resolution (0.37 ml) MRSI within approximately 10 minutes. Moreover, the clean spectra simplify post-processing, enabling the use of peak integration to generate metabolite maps. Our results strongly suggest that downfield MRSI is a promising field for future *in vivo* studies of the brain and, with its short acquisition time, may also be applicable in clinical routine.

## Acknowledgements

The funding was provided by Schweizerischer Nationalfonds zur Förderung der Wissenschaftlichen Forschung, grant numbers: SNSF-182569 and SNSF-207997.

## Authorship contribution statement

Ethics committee approval (JS). Study design and preparation (GW, PR, JS). MRSI methodology (GW, SS, JS). Recruitment of volunteers and patients (AH, PR, IZ, JK, PR). Data analysis (GW, PR, JS). Manuscript preparation (GW, PR, JS). Manuscript editing (all authors). GW and PR contributed equally to the manuscript.

## Declaration of competing interest

Guodong Weng and Johannes Slotboom disclose that the application of CHEAP and SLOW-editing described in the paper has been filed at the International Bureau of WIPO as a PCT patent application.

Patent applicant: Universität Bern.

Status of application: published (WO 2022/229728).

Inventors: Guodong Weng and Johannes Slotboom.

## Data and code availability

All source codes for RF-pulse computation and anonymized MRSI data, with the exception of the clinical MRSI data, are available upon request. Interested parties are encouraged to contact the first author directly to gain access to these resources. The clinical MRSI data cannot be shared due to privacy and ethical restrictions.

## Ethics

This study was approved by the local ethical committee of Bern, Switzerland (Number: NCT04233788). The measurements were performed in conformity with the declaration of Helsinki and all local and national ethical regulations. All subjects and patients signed general consent forms and study-specific consent forms in advance before the scans took place.

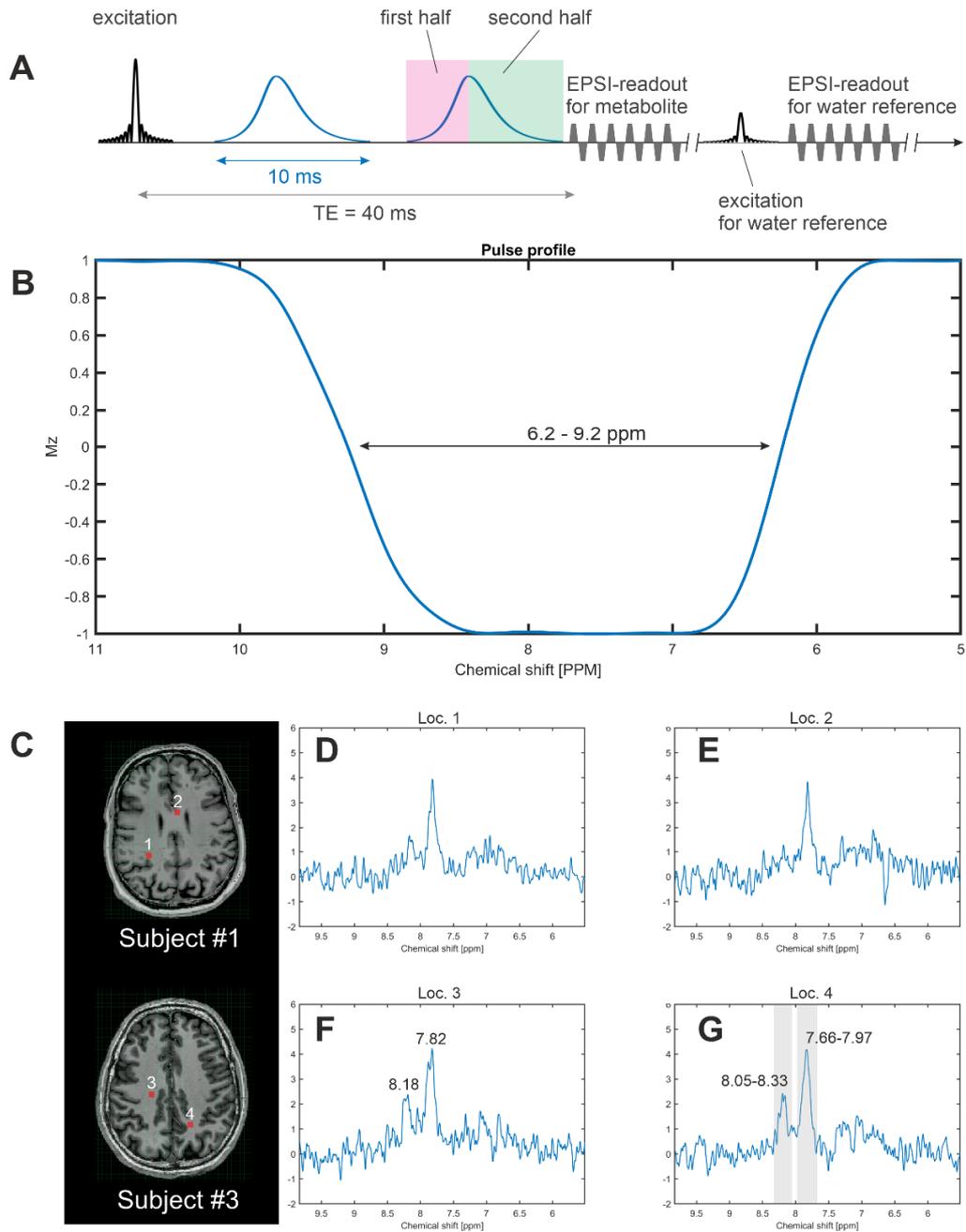

**Figure 1**: **A)** CHEAP-EPSI sequences which makes use of phase compensated adiabatic 2π-pulses with complex secant hyperbolic modulation type (so called CHEAP) [35]. **B)** Pulse profiles of asymmetric CHEAP. **C)** the T1-weighted MRI of subject #1-2. **D)** the spectrum of a selected single voxel (indicated by read area in c) of subject #1. **E)** the spectrum of a selected single voxel (indicated by read area in c) of subject #3. **F)** the spectrum of a selected single voxel (indicated by read area in c) of subject #1. **G)** the spectrum of a selected single voxel (indicated by read area in c) of subject #3.

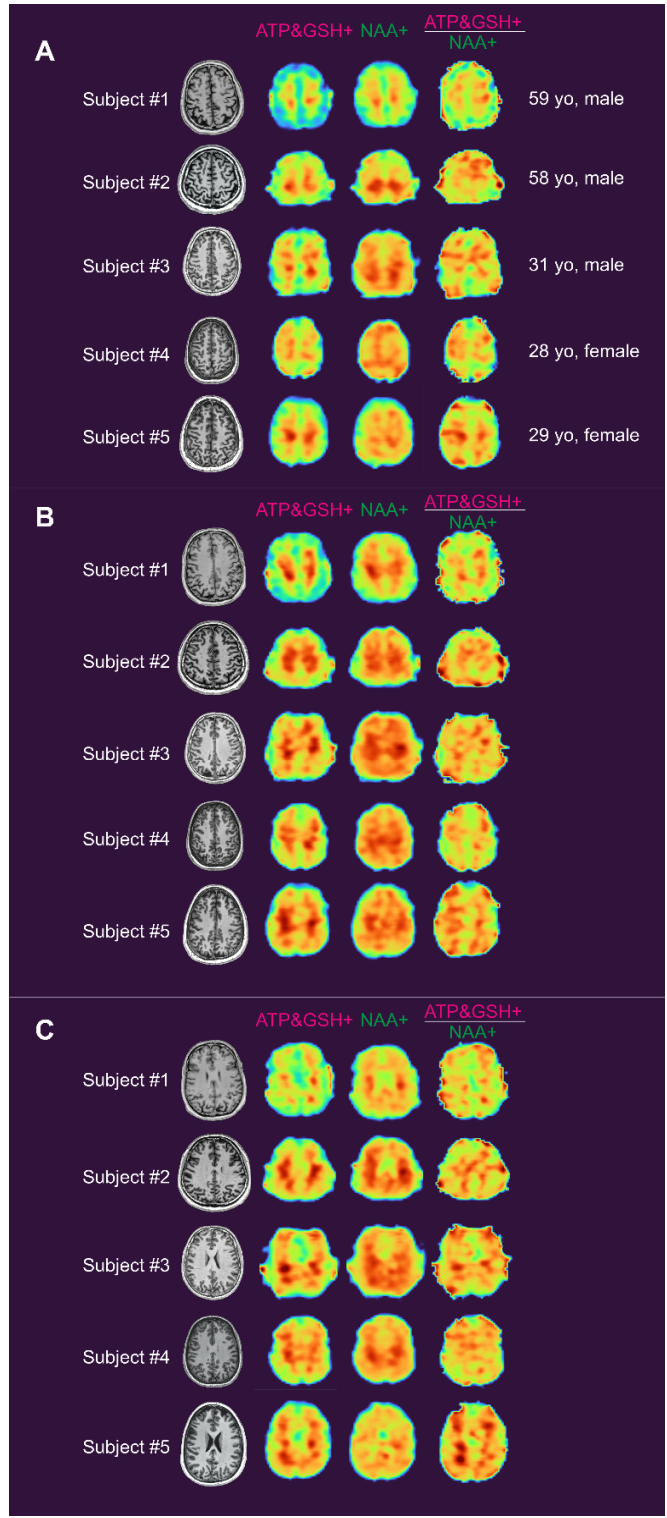

**Figure 2**: The DF 8.18 (ATP&GSH+), 7.82 (NAA+), and 8.18/7.82 peaks integration maps of subject #1-5 with 10 averages (TA = 21 min).

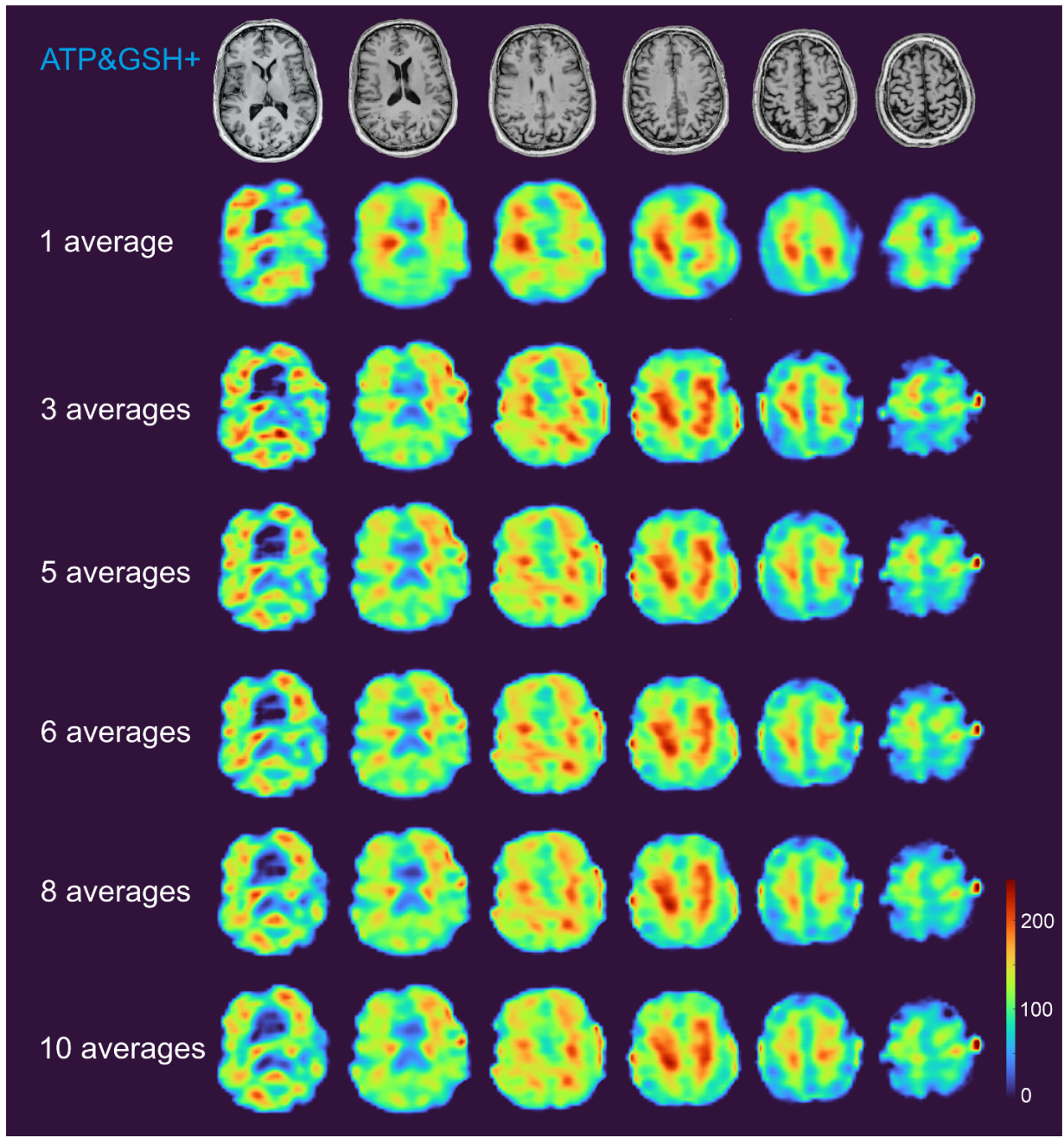

**Figure 3**: The DF 8.18 (ATP&GSH+) peak integration maps of subject #1. For the instance involving 1 average, a 5x5 moving median filter was employed, while a 3x3 moving filter was applied to the remaining cases.

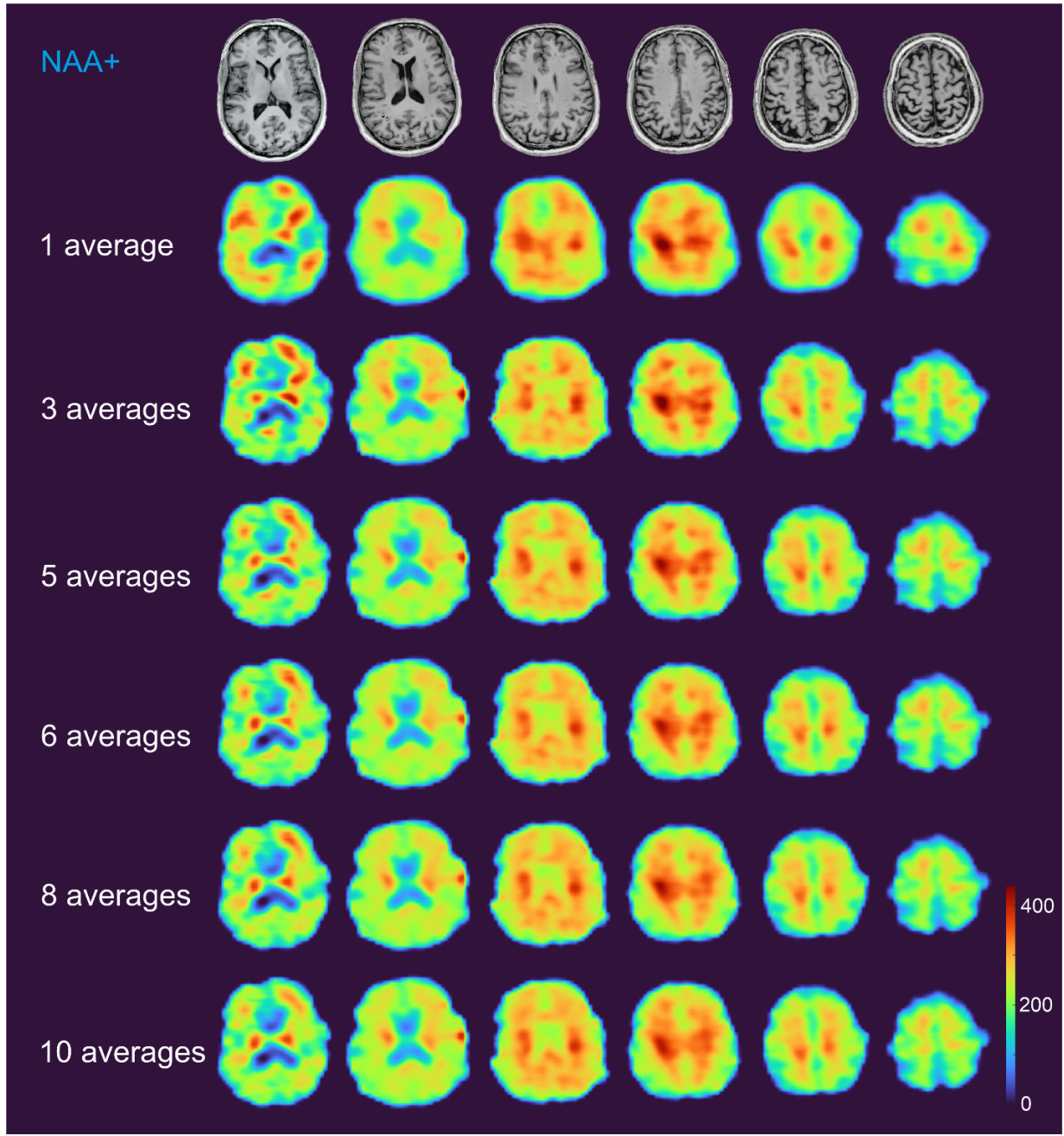

**Figure 4**: The DF 7.82 (NAA+) peak integration maps of subject #1. For the instance involving 1 average, a 5x5 moving median filter was employed, while a 3x3 moving filter was applied to the remaining cases.

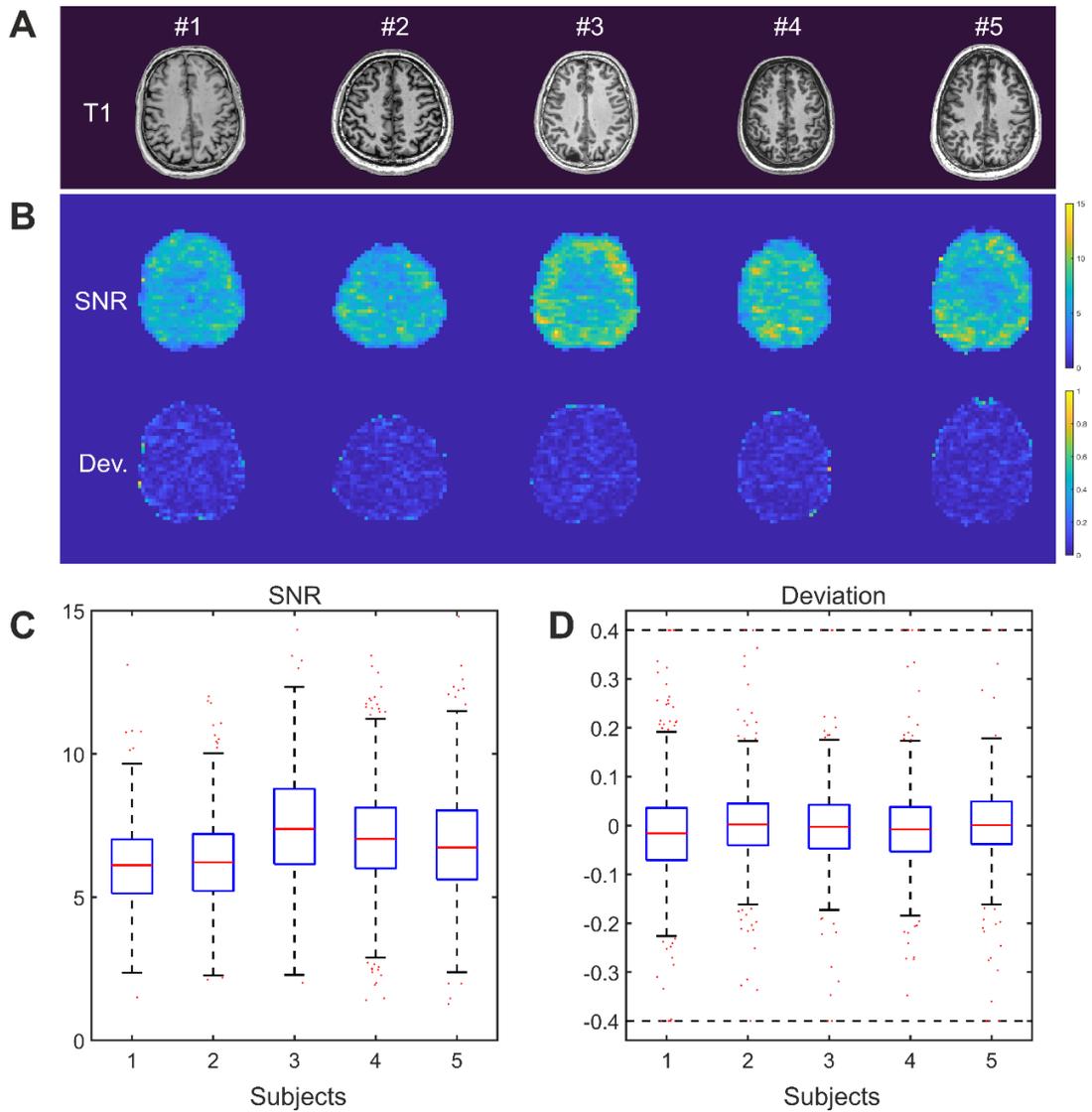

**Figure 5**: SNR and deviation for a selected slice. **A)** T1 for subjects #1-5. **B)** SNR of NAA+ maps with an acquisition of 10.5 minutes; Deviation (absolute value) of NAA+ value between the acquisition of 10.5 and 21 minutes. No filter was applied and displayed voxel size = 4.3×4.3×9.2 mm (nominal voxel size = 4.3×7.8×11 mm = 0.37 ml). **C)** Bar plot for the SNR of NAA+ in B. **D)** Bar plot for the deviation (real value) of NAA+ in B.

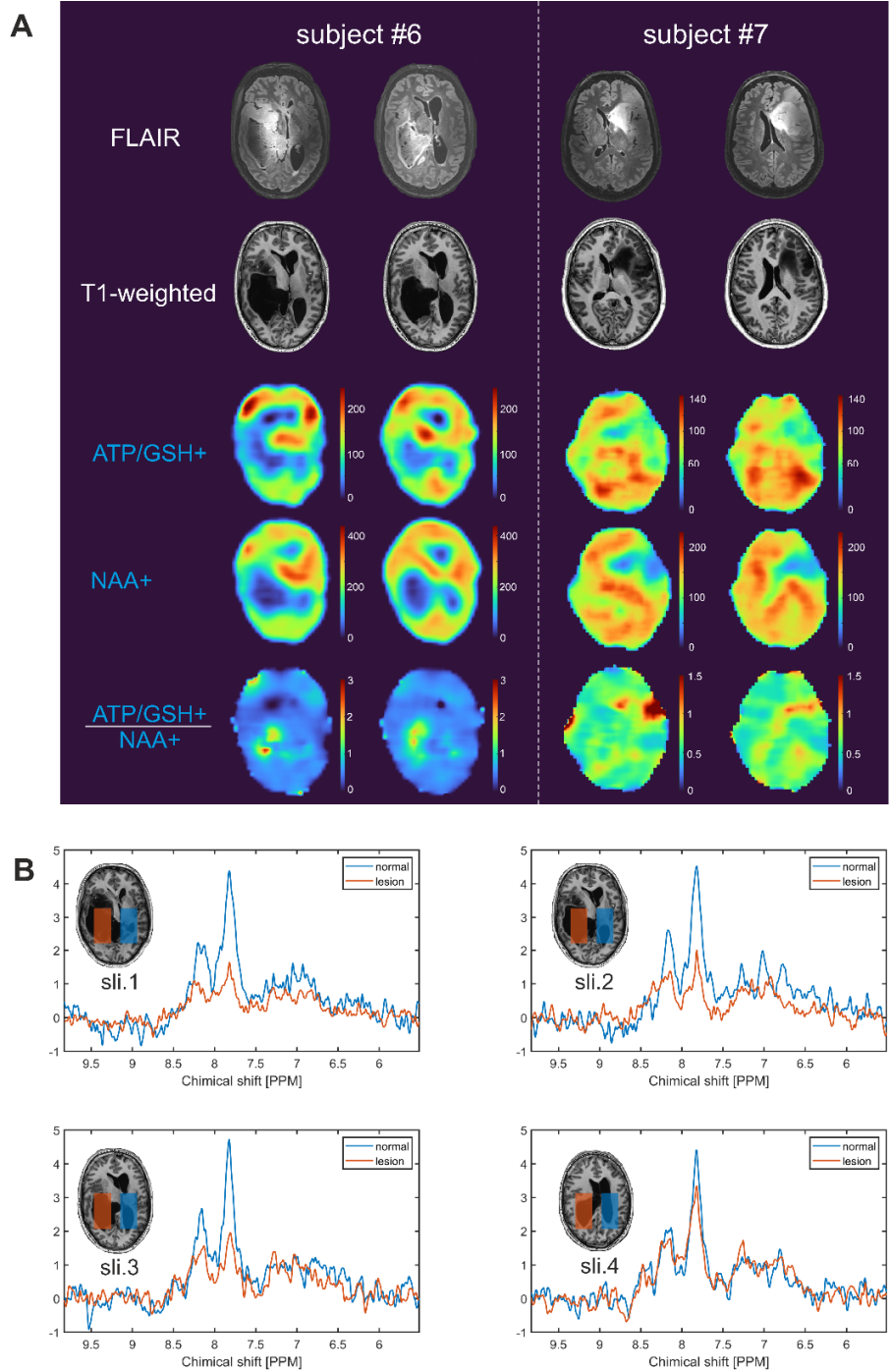

**Figure 6**: Downfield MRSI for glioma patients (subject #6-7). **A)** The DF 8.18, 7.82 (NAA+), and 8.18/7.82 peaks integration maps with 2 averages (TA = 4:12 min) and a 5x5 moving mean filter was applied. **B)** The spectra of lesion and normal tissue in four different slices for subject #5. The selected area size is around 38.8 ml.

# CHEmical-shift selective Adiabatic Pulse (CHEAP): Fast and High Resolution Downfield 3D 1H-MRSI at 7T

## Supplementary material

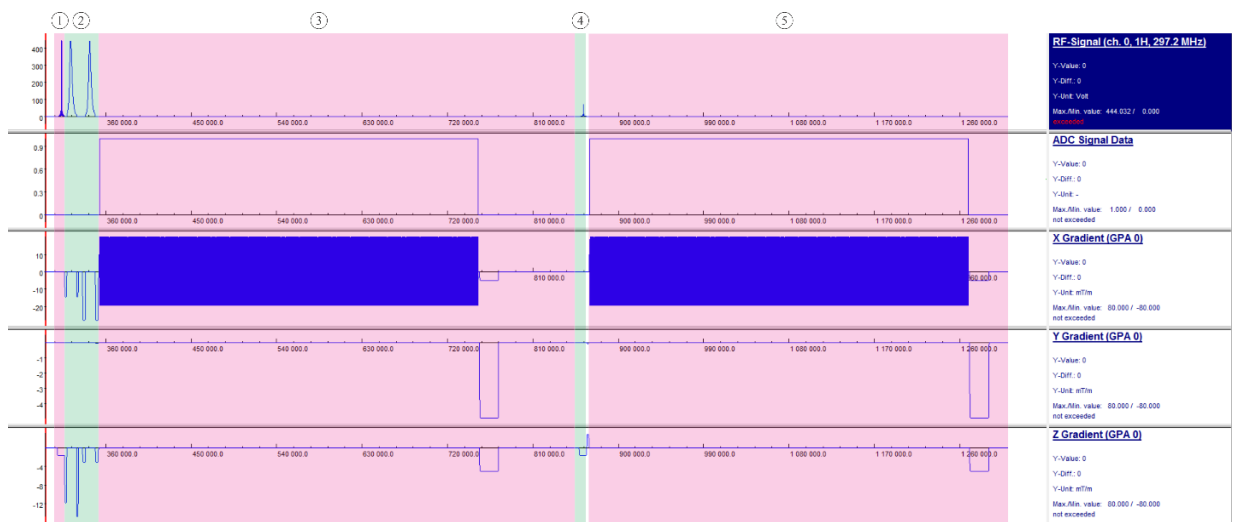

**Figure S1: Sequence scheme in Siemens IDEA VE12U platform.** 1) Slice selective excitation pulse. 2) CHEmical-shift selective Adiabatic Pulse (CHEAP) pair. 3) EPSI metabolite readout. 4) Slice selective excitation pulse for water reference. 5) EPSI water reference readout.

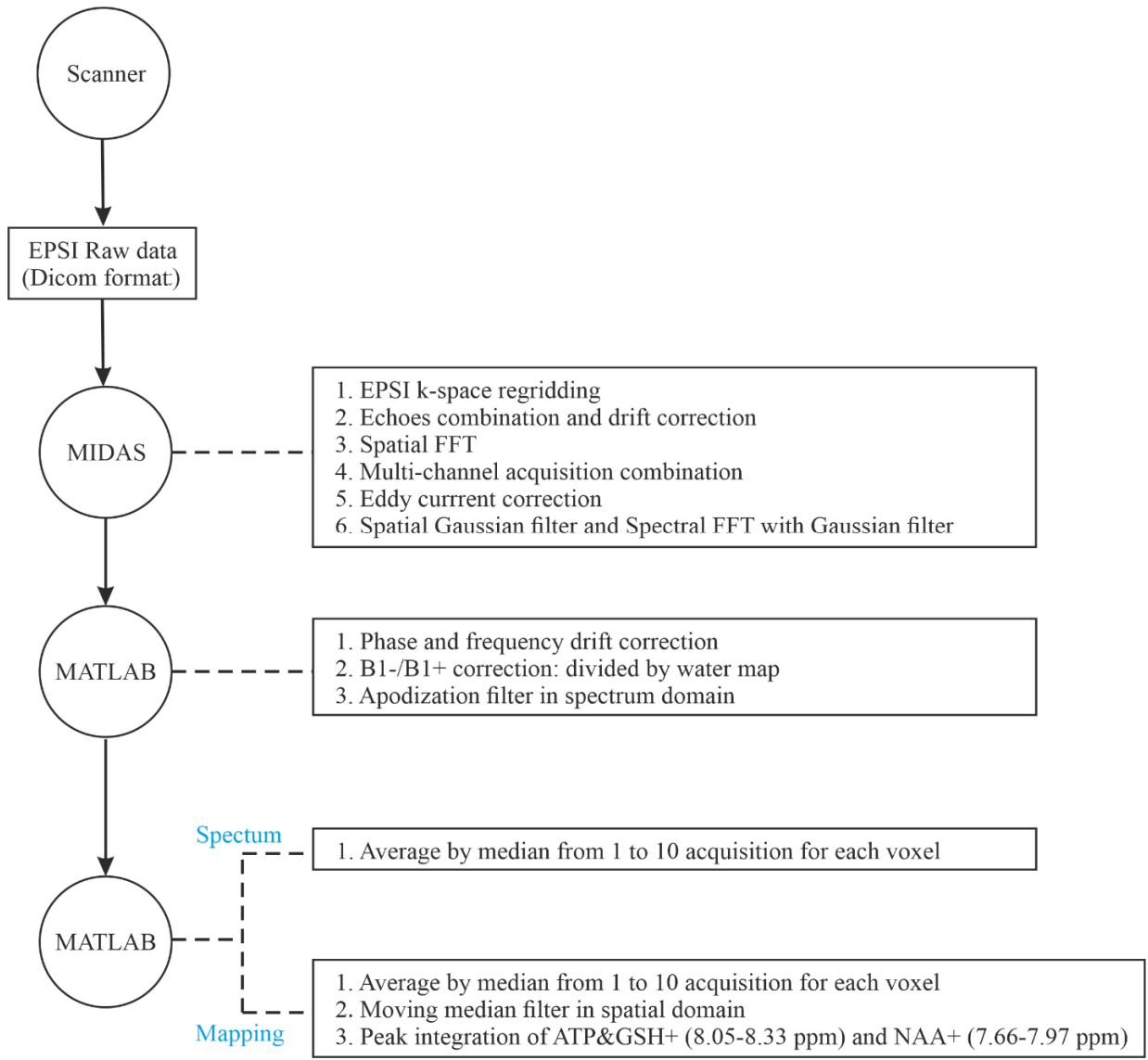

**Figure S2 | The flowchart of reconstruction and processing SLOW-EPSI data.**

The details of each processing step of the raw EPSI data were described in a previous publication [1]. $B_1^-/B_1^+$ correction was performed using water reference data: Meta_corr (spectrum) = Meta (spectrum) / Water (peak area integral).

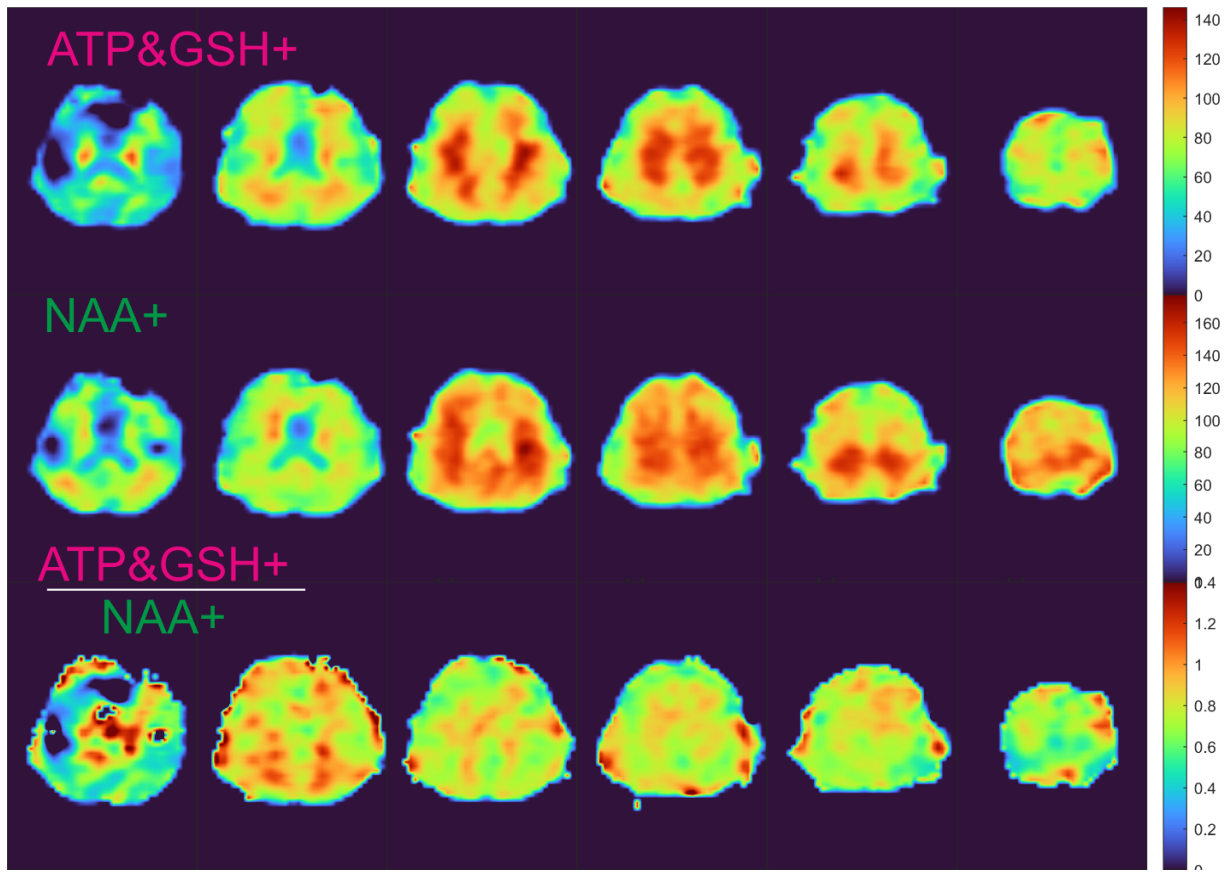

**Figure S3 | Downfield maps for subject #2.** The DF 8.18 (ATP/GSH+), 7.82 (NAA+), and 8.18/7.82 peaks integration maps with 10 averages (TA = 21 min) and a 3x3 moving mean filter was applied

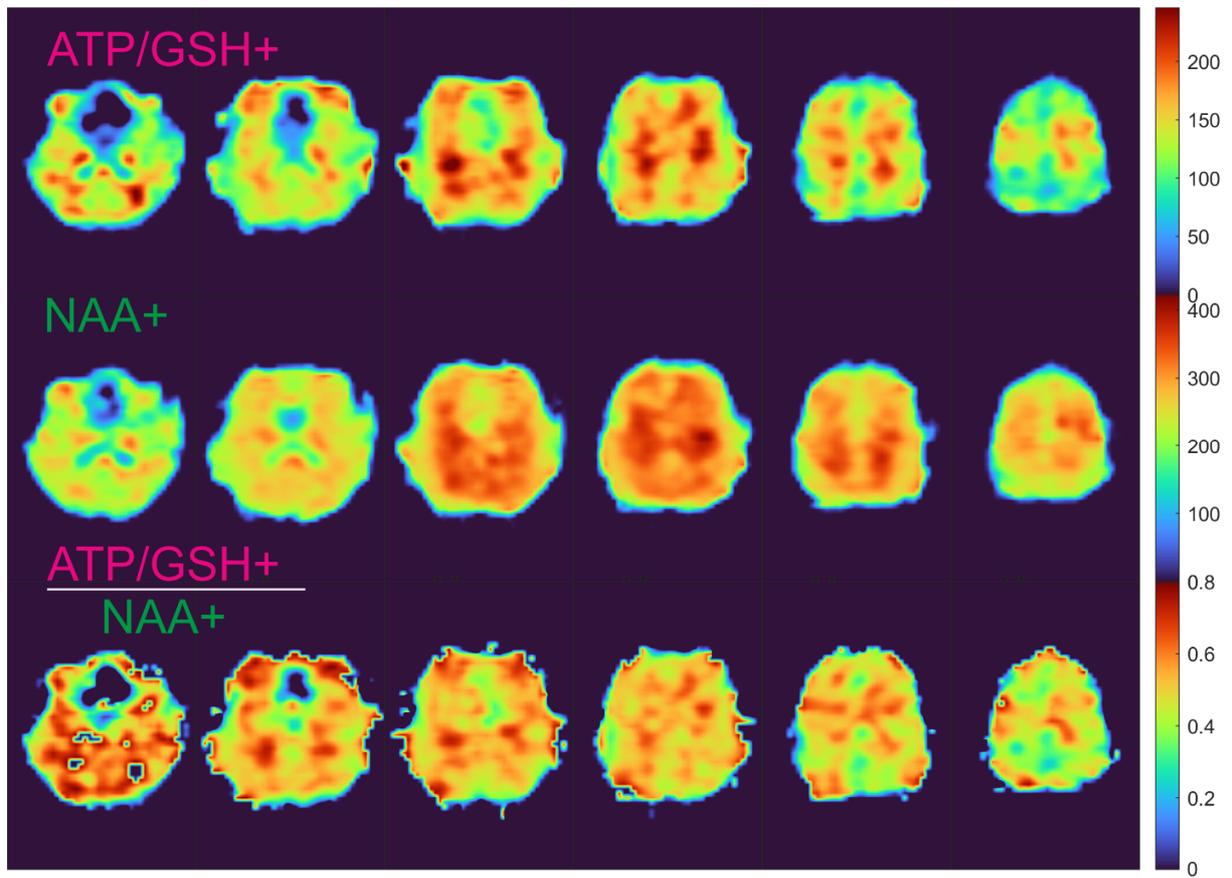

**Figure S4 | Downfield maps for subject #3.** The DF 8.18 (ATP/GSH+), 7.82 (NAA+), and 8.18/7.82 peaks integration maps with 10 averages (TA = 21 min) and a 3x3 moving mean filter was applied.

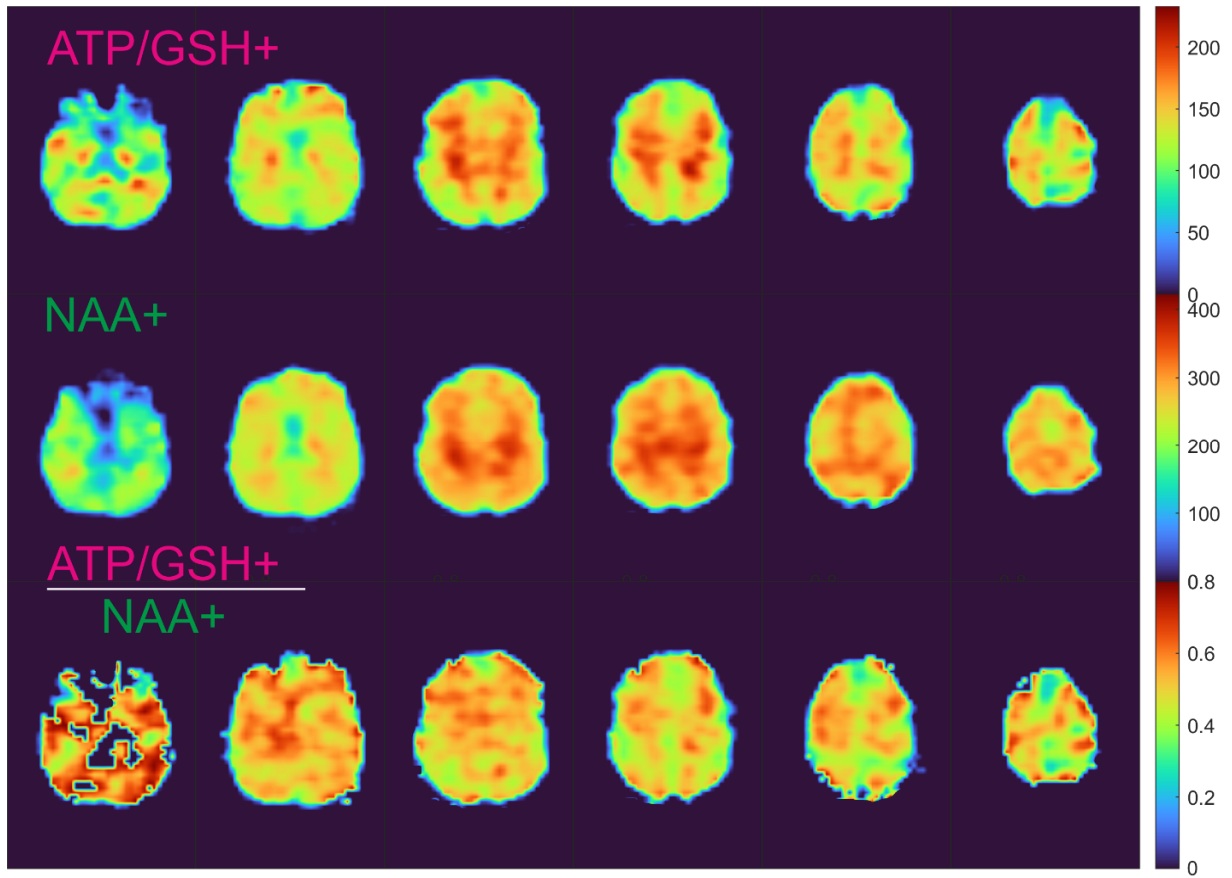

**Figure S5 | Downfield maps for subject #4.** The DF 8.18 (ATP/GSH+), 7.82 (NAA+), and 8.18/7.82 peaks integration maps with 10 averages (TA = 21 min) and a 3x3 moving mean filter was applied.

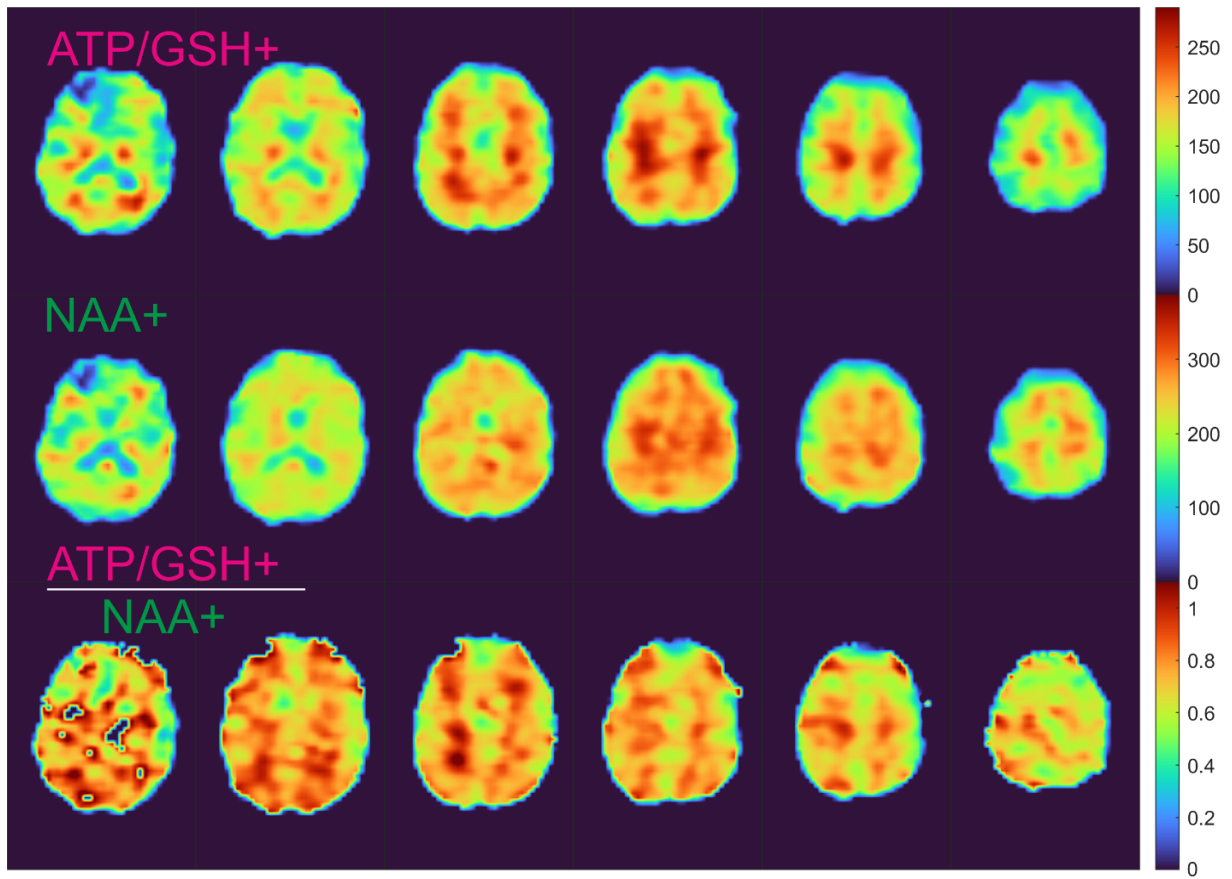

**Figure S6 | Downfield maps for subject #5.** The DF 8.18 (ATP/GSH+), 7.82 (NAA+), and 8.18/7.82 peaks integration maps with 10 averages (TA = 21 min) and a 3x3 moving mean filter was applied.

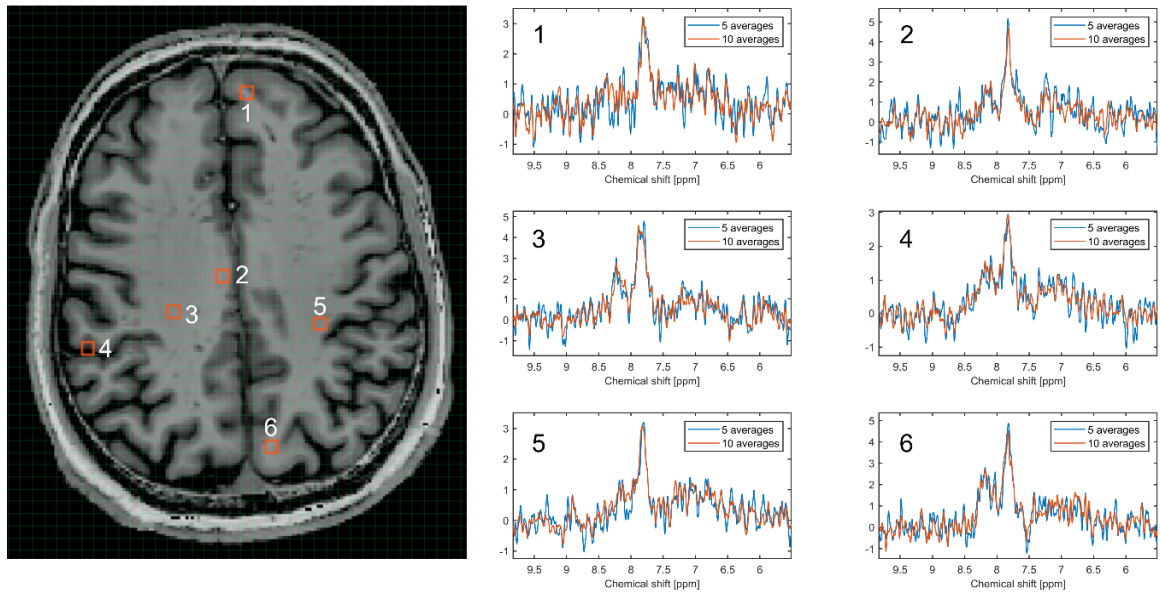

**Figure S7 | Spectra of two acquisitions (10.5 and 21 minutes) of subject #1.** The spectra of 10.5 minutes acquisition are marked as blue, and the spectra of 21 minutes acquisition are marked as orange. Displayed voxel size = 4.3×4.3×9.2 mm (nominal voxel size = 4.3×7.8×11 mm = 0.37 ml).

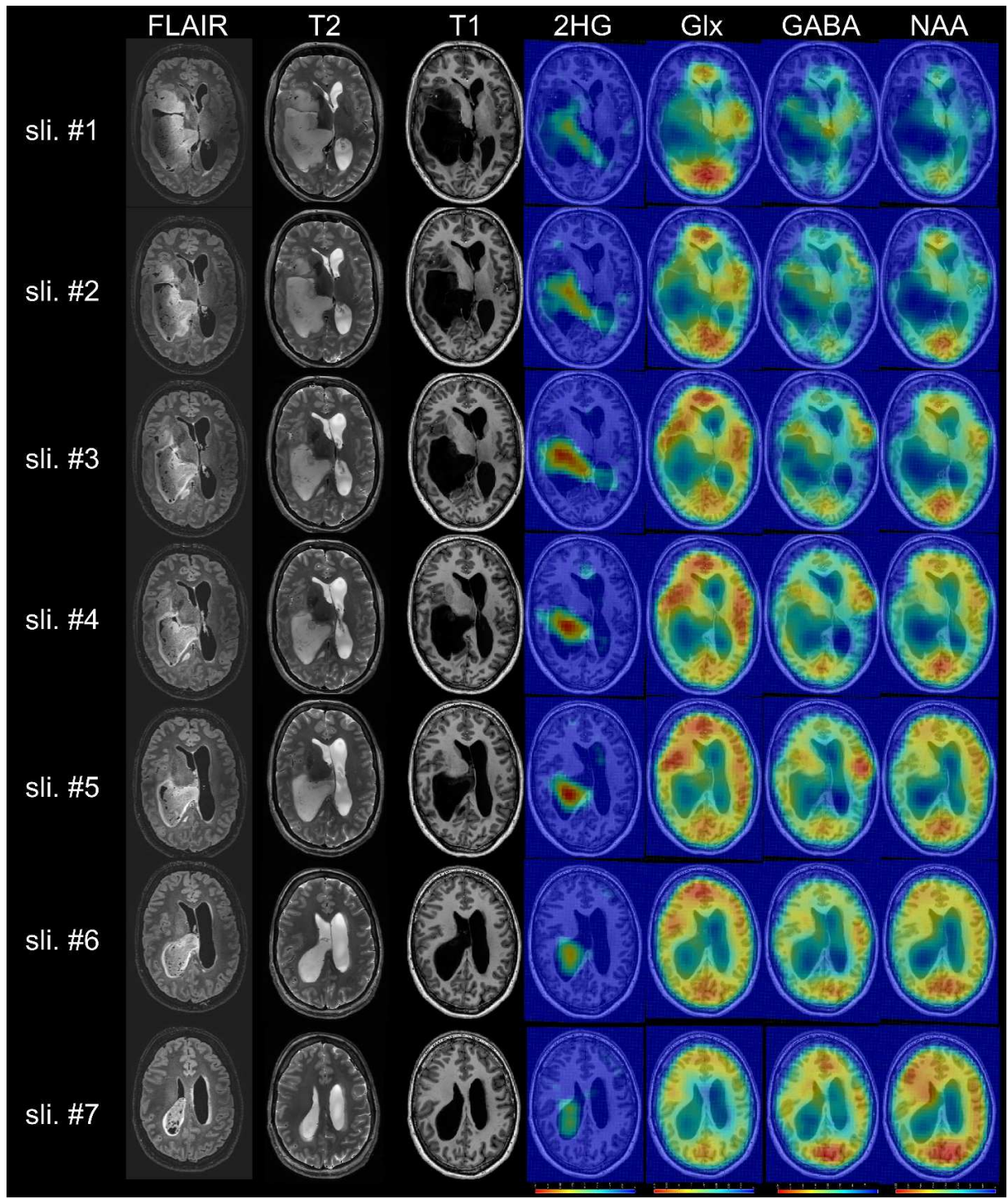

**Figure S8: Upfield and spectral editing MRSI of a glioma patient (subject #6).** A 7×7×3 moving mean filter was applied. TA = 9:04 min with nominal resolution = 4.3×7.8×6.1 mm.